\documentclass[12pt,letterpaper,superscriptaddress,floatfix]{revtex4}
\usepackage[draft]{hyperref}

\usepackage{graphicx}
\usepackage{amsmath}

\begin{document}

\title{Coherent Optical Ultrasound Detection with Rare-Earth Ion Dopants}
\author{Jian Wei Tay}
\author{Patrick M. Ledingham}
\author{Jevon J. Longdell}
\email{jevon@physics.otago.ac.nz}
\affiliation{Jack Dodd Centre, University of Otago, Department of Physics, 730 Cumberland Street, Dunedin, New Zealand}

\begin{abstract}
We describe theoretical and experimental demonstration for optical detection of ultrasound using a spectral hole engraved in cryogenically cooled rare-earth ion doped solids. Our method utilizes the dispersion effects due to the spectral hole to perform phase to amplitude modulation conversion. Like previous approaches using spectral holes it has the advantage of detection with large \textit{\'etendue}. The method also has the benefit that high sensitivity can be obtained with moderate absorption contrast for the spectral holes.
\end{abstract}

\maketitle


Ultrasound is widely used in imaging and non-destructive testing. Most of this testing require an ultrasonic transducer in intimate physical contact with the subject. However, the ability to remotely detect ultrasound removes this restriction allowing a greater variety of applications. A form of remote detection known as air coupled ultrasound \cite{Stoessel02} is under active development. However sensitivity is limited by small coupling between vibrations in the test object and air due to a large impedance mismatch. Remote optical detection uses the fact that light scattered from a vibrating object becomes phase modulated. The challenge of this method is the detection of the phase modulation sidebands on the received light which in general is highly spatially multimode.

For situations where the received light is in a well defined spatial mode, heterodyne detection and other forms of interferometry \cite{dewhurst,scruby} lend themselves readily to the task. Quantum noise limited performance can be relatively easily obtained \cite{bachor}. However in practical applications where the received light has a complex speckle pattern, the interferometric visibility becomes severely affected resulting in low sensitivity.

The ability of an optical system to process light is measured by its \textit{\'etendue} $\mathcal{E} = \mathcal{S} \Omega$, which is the product of the detection surface area $\mathcal{S}$ and the scattering solid angle $\Omega$. Adding optics to the system can change both the collection area and the solid angle but by the constant brightness theorem cannot improve the \'etendue. Interferometry suffers from lower sensitivity as the \'etendue is limited to $\lambda^2$ due to mode-matching requirements \cite{antenna}. Confocal Fabry-Perot resonators can achieve better with \'etendue of around $10^{-3}$ sr mm$^{2}$ \cite{cfpi}.

Adaptive imaging techniques based on photorefractive crystals allow much greater \'etendue compared to classical interferometers \cite{prcrystals,prinvivo}. In these approaches, the photorefractive crystals act as an adaptive beamsplitter which matches the mode of the probe beam travelling through the sample with a local oscillator. Recent advances have demonstrated fast, efficient photorefractive crystals able to operate in the ``therapeutic window''\cite{farahi}.
However the programming speed utilized for imaging demonstrations ($\tau_{PR}=100~ms$) are still slow for biological imaging of thick tissue \textit{in vivo}, where speckle decorrelation times are on the order of $0.1~ms$ \cite{gross}.

An adaptive detection scheme using rubidium vapor was recently proposed and demonstrated \cite{rbholography}. This scheme has the advantage of shorter writing times compared to photorefractive crystals. The scheme employs the use of self-rotation of polarized light to ``program'' the hologram in the atomic vapour. However, it is unclear how the sensitivity of the detection relies on the angular dependence of the beams. In \cite{rbholography}, near parallel signal and reference beams were used which means there is little Doppler broadening of the two-photon transitions. It is unclear how effective the detection will be when the beams are more than a few degrees apart as Doppler broadening in the hyperfine transitions used will affect the polarization self-rotation \cite{rochester}.

The suitability of rare-earth ion doped crystals as spectral filters to detect ultrasound has recently been demonstrated by Li \textit{et al.} \cite{li08}. In that work the received light was directed through a Tm:YAG crystal that had a spectral hole prepared at the same frequency as one of the modulation sidebands. The sample absorbed the unmodulated light but let the sideband through, allowing it to be measured directly. 

Rare-earth doped crystals used in this manner allow \'etendue several orders of magnitude higher than is able to be achieved using other techniques. However detecting weak ultrasound signals, where the unmodulated component is much larger than the modulated component, using this approach requires large absorption away from the hole while simultaneously obtaining high transmission through a narrow hole. Although very high contrast spectral filtering has been demonstrated for Pr$^{3+}$ \cite{hedges}, the transition of interest in this case is at 606~nm where laser sources are difficult to obtain. For other dopants with more convenient wavelengths, such as Tm$^{3+}$, shorter hole lifetimes makes this much more difficult.

Here we demonstrate an alternate scheme using the dispersive properties of an engraved spectral hole. Our method utilizes Tm$^{3+}$ ions which have resonance at 793~nm and are able to be probed using diode and other solid state lasers. We demonstrate that high sensitivity can be achieved using sensible hole burning parameters.

Our technique relies on the steep dispersion of the spectral hole to convert phase modulation into amplitude modulation.  By applying a phase shift to the carrier that is different to that of the two sidebands, phase modulation can be transformed efficiently into amplitude modulation. Using the carrier as a local oscillator allows shot noise limited detection of this amplitude modulation. The technique has a lot in common with Pound-Drever-Hall laser stabilization \cite{dreverhall} to a spectral hole. However in this case we are deliberately offset a fixed amount from hole center and are measuring the phase modulation rather than the laser frequency.

We now provide the theoretical description of our system. Consider a monochromatic laser beam reflected off the surface of an object vibrating at ultrasonic frequency $\omega_M$. The reflected light\cite{bjorklund,supplee} can be described as

\begin{equation}
E_{r}(t) = E_0 \exp(\omega_c) \exp(-i M\,{\rm Re} \{\exp (i\omega_M t)\}) ,
\end{equation}
where $\omega_c$ is the optical frequency, $E_0$ is the amplitude of the electric field and $M$ is the (assumed real) modulation index $M = 4\pi U/\lambda$, where $U$ is the amplitude of the ultrasonic displacement and $\lambda$ is the optical wavelength. For ultrasonic displacements small compared to the wavelength of light (i.e. $M\ll 1$) this becomes
\begin{equation}
E_{r}(t) = E_0  \left[-i \frac{M}{2} \exp(-i \omega_M t)  + 1 + i \frac{M}{2} \exp(i \omega_M t)\right] .
\end{equation}

In our experiment, we used the $^3H_6 \rightarrow\,^3H_4$ transition in Tm:YAG which occurs around 793~nm. The sample used had an inhomogeneous linewidth of around 30~GHz. This large inhomogeneous broadening means we can assume the absorption spectrum is flat except where the ions have been optically pumped into the metastable (10~ms) shelving state\ $^3F_4$, resulting in a spectral hole.

We can assume that the Tm:YAG crystal so prepared has a complex transmission function $T(\omega)$. The phasor for the light transmitted through the crystal is therefore given by
\begin{eqnarray}
\nonumber
E_{r}(t) = E_0  \left[-i \frac{M}{2} T(-\omega_M) \exp(-i \omega_M t)  + T(0) \right.&\\
\left.  + i \frac{M}{2} T(\omega_M) \exp(i \omega_M t)\right] &.
\end{eqnarray}

The detected photocurrent that results is given by  $I(t) =\eta q /(\hbar \omega_c) (\epsilon_0 c A/2)\, E_r(t)^\ast\cdot E_r(t)$. Hence we obtain

\begin{equation}
  I(t) = I_0(1 + M Re\{ \beta \exp(i\omega_mt)\}) ,\label{eq:mod-xfer}
\end{equation}
where $\beta = T(\omega_m)T(0)^* - T(-\omega_m)^*T(0)$.

In order to achieve shot noise limited sensitivity two things are required: firstly there can be no loss from the sidebands. Secondly, as described by Eq.~(\ref{eq:mod-xfer}), the carrier needs to be phase shifted by $\pi/2$ relative to the mean phase of the two sidebands in order to transform the phase modulation completely into amplitude modulation. These could be achieved by having broad spectral holes for the sidebands and an absorptive spike, such as that used in \cite{tomog,hedges}, sitting to one side of the carrier.

Here we use a simpler absorption profile using a spectral hole offset from the carrier. The complex transmission function for the spectral hole can be modelled by \cite{bottger,julsgaard}

\begin{equation}
  T(\omega) = \exp \left[ - \frac{\alpha L}{2} \left(1 - \frac{\Gamma}{\Gamma - i \Delta }\right) \right] ,
  \label{eq:ctf}
\end{equation}
where $\alpha L$ is the optical depth, $\Gamma/2 \pi$ is the hole linewidth and $\Delta$ is the angular frequency offset between the laser frequency and the center of the hole. When the laser frequency is offset from the hole center, broadband detection is obtained so long as the modulation frequency is larger than the hole width.

%
\begin{figure}[h]
\includegraphics{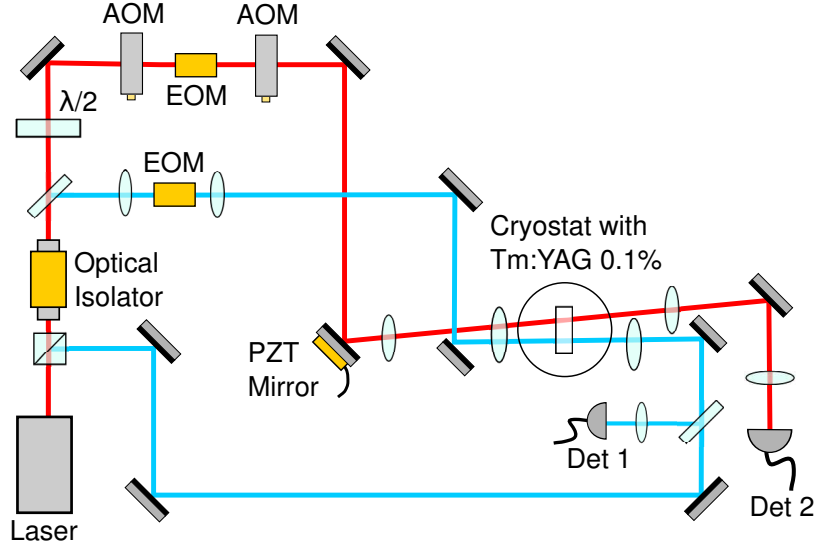}
\caption{(Color online) Experimental setup. The experimental beam (in red) is frequency shifted and gated using two AOMs. The ultrasonic pulses were applied using a PZT-backed mirror. The beam is then steered through a prepared Tm:YAG 0.1\% crystal before being detected using a photodetector. The laser is locked to a spectral hole using a method similar to Bottger \textit{et al.} \cite{bottger} (path in blue), except with optical feedback \cite{tobepublished}. \label{fig:Experiment}}
\end{figure}

Our experimental setup is shown in Figure~\ref{fig:Experiment}, with the red beam indicating the experimental beam path and the blue indicating the locking beam path. We used an extended cavity diode laser operating at 793 nm, frequency stabilized using a hybrid optical and electronic technique \cite{tobepublished}. The laser is locked to a spectral hole engraved in the same crystal providing a narrow spectral filter for the seeding beam. This gives a dramatically narrower Schawlow-Townes linewidth and much reduced sensitivity to current noise. We measure an 30~dB drop in the amount of phase noise around 1~MHz.

To perform ultrasound measurements, the beam frequency is shifted and gated using two AOMs. The ultrasound pulses were generated using either an EOM for calibration of the detection system, or a PZT backed mirror operating at 1.11 MHz. The modulated beam is then directed through a rare-earth ion doped crystal. The sample used was 0.1\%~Tm$^{3+}$:YAG (Scientific Materials) cryogenically cooled to 2.9K. The incident beam on the sample had a power of 0.6 mW and a diameter of 1 mm. The beam arriving at the photodetector had a power of 146 ${\rm \mu}$W. 

To generate the signal, the output from the photodetector was AC coupled then amplified using a low noise RF amplifier (Minicircuits ZFL-500LN). The amplified signal was then downconverted by mixing with a local oscillator at the same RF frequency as the applied ultrasound modulation. This was then amplified using a preamplifier with a gain of 100 and a low pass filter set to a 30~kHz bandwidth before measurement with an oscilloscope.

\begin{figure}[ht]
\includegraphics{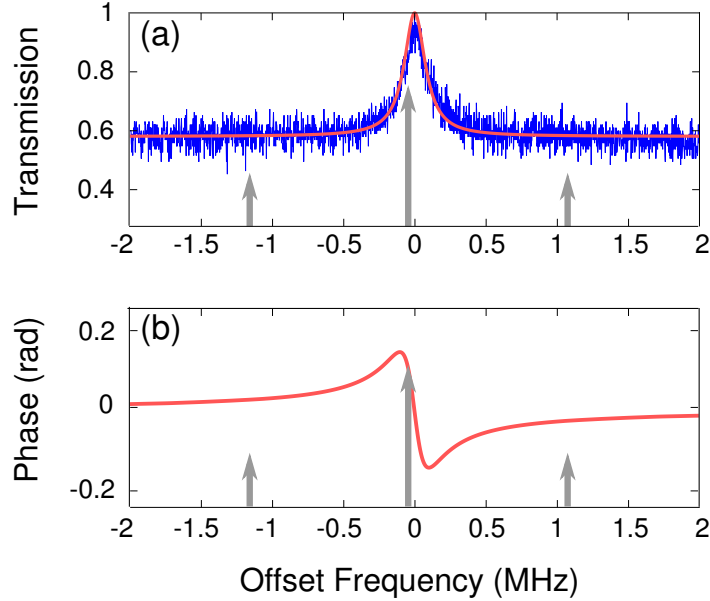}
\caption{(Color online) Response of a spectral hole. This shows (a) the normalized transmission through the hole and its calculated fit (in red), and (b) the phase response as given by the Kramers-Kr\"onig relations. The arrows indicate the position of the sideband and carrier frequencies of the ultrasonically modulated light. \label{fig:holeResponse}}
\end{figure}

The properties of the engraved spectral hole is shown in Figure~\ref{fig:holeResponse}. The (normalized) transmission is measured by sweeping the laser frequency 4 MHz across the hole and is shown in subplot (a). Using Equation~\ref{eq:ctf}, we can fit to the trace to obtain an optical depth $\alpha L$ of 0.5 and linewidth $\frac{\Gamma}{2 \pi}$ of 371~kHz. The phase shift was then calculated using the Kramers-Kr\"onig relations and is shown in Figure~\ref{fig:holeResponse} (b). For the optical depth used, the maximum phase shift obtained is around 0.14 rad compared to the ideal case of $\pi/2$.

\begin{figure}[h]
\includegraphics[width=11cm]{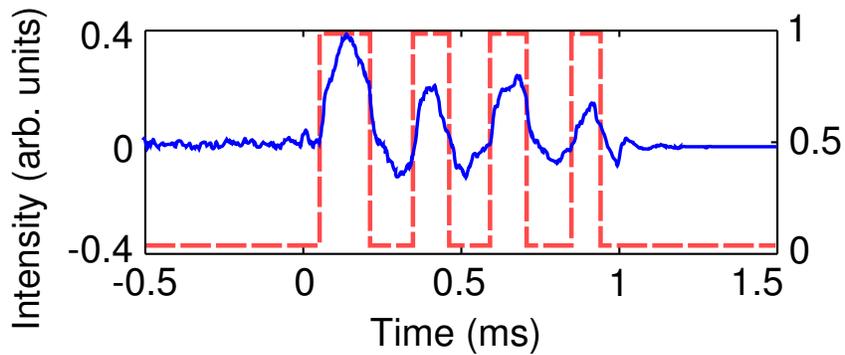}
\caption{(Color online) Ultrasonic pulse detection. The detected intensity (blue solid line) is shown with respect to the applied pulses (red dashed line). The axis on the right indicates the normalized pulse amplitude. \label{fig:Pulsed}}

\end{figure}

We now demonstrate the detection of ultrasound pulses using a spectral hole engraved as before. The laser frequency is shifted by 39 kHz to sit the laser to the side of the spectral hole as indicated in Figure~\ref{fig:holeResponse} in order to obtain a large phase shift. Ultrasound pulses were then applied using the PZT backed mirror and the retrieved signal is shown in Figure~\ref{fig:Pulsed}.

We can calculate the position equivalent noise on our system from the ratio of detection noise with the sensitivity afforded by the spectral hole. Assuming shot noise and an optimum optical depth such that $|\beta| = 1$, the optimal position equivalent noise for this technique was calculated to be $1.3 \times 10^{-13}$ m/$\sqrt{\rm Hz}$. However, in practice the position equivalent noise on our system was  measured to be $4.6 \times 10^{-11}$ m /$\sqrt{\rm Hz}$ given a 30 kHz detection bandwidth. The reason for increased noise on our detection is two-fold: the sample used has a non-optimum optical depth leading to reduced phase-to-amplitude conversion, and there is residual phase noise on the laser at ultrasonic frequencies.

In conclusion, we have demonstrated sensitive coherent optical detection of ultrasound using the dispersive properties of spectral holes. The achieved position equivalent noise is approximately one hundred times larger than ideal shot noise limited detection. There are two reasons for this firstly the conversion of phase to amplitude modulation with the non-ideal spectral hole shape results in a sensitivity one fifth of the ideal case. Excess noise on the laser above the shot noise level accounts for the rest. The technique is suitable for the optical detection ultrasound with large \'etendue and in principle could reach shot noise limited sensitivities with spectral hole-burning materials of modest performance.

This work was produced with the support of the New Zealand Foundation for Research, Science and Technology under Contract No. NERF-UOOX0703.

\end{document}